\documentclass[a4paper]{jpconf}
\usepackage[centertags]{amsmath}
\usepackage{amsbsy}
\usepackage{amsfonts}
\usepackage{amssymb}
\usepackage{graphicx}
\usepackage{bm}
\usepackage{cite}
\newcommand{\nua}[1]{\ensuremath{\rlap
           {\kern-2.5pt\ensuremath
           {\overset{\scriptscriptstyle(-)}{\phantom{\nu}}}}
           {\ensuremath{{\nu}_{#1}}}}}
\begin{document}

\title{Sterile Neutrino Status}

\author{Carlo Giunti}

\address{INFN, Sezione di Torino, Via P. Giuria 1, I--10125 Torino, Italy}

\ead{giunti@to.infn.it}

\begin{abstract}
We review the results of global analyses of short-baseline neutrino oscillation data
in
3+1,
3+2 and
3+1+1
neutrino mixing schemes.
\\
\textbf{Talk presented at NuFact 2013, 15th International Workshop on Neutrino Factories, Super Beams and Beta Beams, 19-24 August 2013, IHEP, Beijing, China.}
\end{abstract}

\section{Introduction}
\label{Introduction}

Neutrino oscillations have been measured with high accuracy in
solar, atmospheric and long-baseline
neutrino oscillation experiments
(see \cite{1205.4018,1205.5254,1209.3023}).
Hence, we know that
neutrinos are massive and mixed particles
(see \cite{Giunti-Kim-2007,0704.1800})
and there are two independent squared-mass differences:
the solar
$\Delta{m}^2_{\text{SOL}} \simeq 7.5 \times 10^{-5} \, \text{eV}^2$
and the atmospheric
$\Delta{m}^2_{\text{ATM}} \simeq 2.3 \times 10^{-3} \, \text{eV}^2$.
This is in agreement with the standard
three-neutrino mixing paradigm,
in which the three active neutrinos
$\nu_{e}$,
$\nu_{\mu}$,
$\nu_{\tau}$
are superpositions of three massive neutrinos
$\nu_1$,
$\nu_2$,
$\nu_3$
with respective masses
$m_1$,
$m_2$,
$m_3$.
The two measured squared-mass differences can be interpreted as
$
\Delta{m}^2_{\text{SOL}}
=
\Delta{m}^2_{21}
$
and
$
\Delta{m}^2_{\text{ATM}}
=
|\Delta{m}^2_{31}|
\simeq
|\Delta{m}^2_{32}|
$,
with
$\Delta{m}^2_{kj}=m_k^2-m_j^2$.

The completeness of the three-neutrino mixing paradigm has been challenged by
the following indications in favor of short-baseline neutrino oscillations,
which require the existence of at least one additional squared-mass difference,
$\Delta{m}^2_{\text{SBL}}$,
which is much larger than
$\Delta{m}^2_{\text{SOL}}$
and
$\Delta{m}^2_{\text{ATM}}$:

\begin{enumerate}

\renewcommand{\labelenumi}{\theenumi.}
\renewcommand{\theenumi}{\arabic{enumi}}

\item
The
LSND experiment,
in which a signal of short-baseline
$\bar\nu_{\mu}\to\bar\nu_{e}$
oscillations has been observed
with a statistical significance of about $3.8\sigma$
\cite{nucl-ex/9504002,hep-ex/0104049}.

\item
The reactor antineutrino anomaly
\cite{1101.2755},
which is a deficit of the rate of $\bar\nu_{e}$ observed in several
short-baseline reactor neutrino experiments
in comparison with that expected from a new calculation of
the reactor neutrino fluxes
\cite{1101.2663,1106.0687}.
The statistical significance is about $2.8\sigma$.

\item
The Gallium neutrino anomaly
\cite{nucl-ex/0512041,Laveder:2007zz,hep-ph/0610352,1006.3244,1210.5715},
consisting in a short-baseline disappearance of $\nu_{e}$
measured in the
Gallium radioactive source experiments
GALLEX
\cite{1001.2731}
and
SAGE
\cite{0901.2200}
with a statistical significance of about $2.9\sigma$.

\end{enumerate}

In this review, we consider
3+1
\cite{hep-ph/9606411,hep-ph/9607372,hep-ph/9903454,hep-ph/0405172},
3+2
\cite{hep-ph/0305255,hep-ph/0609177,0705.0107,0906.1997}
and
3+1+1
\cite{1010.3970,1201.6662,1205.1791,1306.6079}
neutrino mixing schemes
in which there are one or two additional massive neutrinos at the eV scale
and
the masses of the three standard massive neutrinos are much smaller.
Since from the LEP measurement of the invisible width of the $Z$ boson
we know that there are only three active neutrinos
(see \cite{Giunti-Kim-2007}),
in the flavor basis the additional massive neutrinos correspond to
sterile neutrinos
\cite{Pontecorvo:1968fh},
which do not have standard weak interactions.

The possible existence of sterile neutrinos
is very interesting, because they are new particles which could
give us precious information on the physics beyond the Standard Model
(see \cite{hep-ph/0111326,hep-ph/0603118}).
The existence of light sterile neutrinos is also very important for astrophysics
(see \cite{1206.6231})
and cosmology
(see \cite{1301.7102,1307.0637}).

In the 3+1 scheme,
the effective probability of
$\nua{\alpha}\to\nua{\beta}$
transitions in short-baseline experiments has the two-neutrino-like form
\begin{equation}
P_{\nua{\alpha}\to\nua{\beta}}
=
\delta_{\alpha\beta}
-
4 |U_{\alpha4}|^2 \left( \delta_{\alpha\beta} - |U_{\beta4}|^2 \right)
\sin^2\!\left( \dfrac{\Delta{m}^2_{41}L}{4E} \right)
\,,
\label{pab}
\end{equation}
where $U$ is the mixing matrix,
$L$ is the source-detector distance,
$E$ is the neutrino energy and
$\Delta{m}^2_{41} = m_{4}^2 - m_{1}^2 = \Delta{m}^2_{\text{SBL}} \sim 1 \, \text{eV}^2$.
The electron and muon neutrino and antineutrino appearance and disappearance
in short-baseline experiments
depend on
$|U_{e4}|^2$ and $|U_{\mu4}|^2$,
which
determine the amplitude
$\sin^22\vartheta_{e\mu} = 4 |U_{e4}|^2 |U_{\mu4}|^2$
of
$\nua{\mu}\to\nua{e}$
transitions,
the amplitude
$\sin^22\vartheta_{ee} = 4 |U_{e4}|^2 \left( 1 - |U_{e4}|^2 \right)$
of
$\nua{e}$
disappearance,
and
the amplitude
$\sin^22\vartheta_{\mu\mu} = 4 |U_{\mu4}|^2 \left( 1 - |U_{\mu4}|^2 \right)$
of
$\nua{\mu}$
disappearance.

Since the oscillation probabilities of neutrinos and antineutrinos are related by
a complex conjugation of the elements of the mixing matrix
(see \cite{Giunti-Kim-2007}),
the effective probabilities of short-baseline
$\nu_{\mu}\to\nu_{e}$ and $\bar\nu_{\mu}\to\bar\nu_{e}$
transitions are equal.
Hence,
the 3+1 scheme cannot explain a possible CP-violating difference of
$\nu_{\mu}\to\nu_{e}$ and $\bar\nu_{\mu}\to\bar\nu_{e}$
transitions in short-baseline experiments.
In order to allow this possibility,
one must consider a 3+2 scheme,
in which, there are four additional effective mixing parameters in short-baseline experiments:
$\Delta{m}^2_{51}$,
which is conventionally assumed $\geq\Delta{m}^2_{41}$,
$|U_{e5}|^2$, $|U_{\mu5}|^2$
and
$\eta = \text{arg}\left[U_{e4}^*U_{\mu4}U_{e5}U_{\mu5}^*\right]$.
Since this complex phase appears with different signs in
the effective 3+2 probabilities of short-baseline
$\nu_{\mu}\to\nu_{e}$ and $\bar\nu_{\mu}\to\bar\nu_{e}$
transitions, it can generate measurable CP violations.

A puzzling feature of the 3+2 scheme
is that it needs the existence of two sterile neutrinos
with masses at the eV scale.
We think that it may be considered as more plausible that
sterile neutrinos have a hierarchy of masses.
Hence, it is interesting to consider also the 3+1+1 scheme
\cite{1010.3970,1201.6662,1205.1791,1306.6079},
in which $m_{5}$ is much heavier than 1 eV
and the oscillations due to
$\Delta{m}^2_{51}$
are averaged.
Hence,
in the analysis of short-baseline data
in the 3+1+1 scheme
there is one effective parameter less than in the 3+2 scheme
($\Delta{m}^2_{51}$),
but CP violations generated by $\eta$ are observable.

\section{Global Fits}
\label{Global Fits}

Global fits of short-baseline neutrino oscillation data have been presented recently in
Refs.~\cite{1303.3011,1308.5288}.
These analyses take into account the final results of the
MiniBooNE experiment,
which was made in order to check the LSND signal
with about one order of magnitude larger distance ($L$) and energy ($E$),
but the same order of magnitude for the ratio $L/E$
from which neutrino oscillations depend.
Unfortunately, the results of the
MiniBooNE experiment are ambiguous,
because the LSND signal was not seen in neutrino mode
\cite{0812.2243}
and the signal observed in 2010 \cite{1007.1150}
with the first half of the antineutrino data
was not observed in the second half of the data
\cite{1303.2588}.
Moreover,
the MiniBooNE data in both neutrino and antineutrino modes
show an excess in the low-energy bins
which is widely considered to be anomalous
because it is at odds with neutrino oscillations
\cite{1109.4033,1111.1069}\footnote{
The interesting possibility of reconciling the low--energy anomalous data with neutrino oscillations
through energy reconstruction effects proposed in Refs.~\cite{Martini:2012fa,Martini:2012uc}
still needs a detailed study.
}.

In the following we summarize the results of the analysis of short-baseline data
presented in
Ref.~\cite{1308.5288}
of the following three groups of experiments:

\begin{enumerate}

\renewcommand{\labelenumi}{(\theenumi)}
\renewcommand{\theenumi}{\Alph{enumi}}

\item
The
$\nua{\mu}\to\nua{e}$
appearance data of the
LSND \cite{hep-ex/0104049},
MiniBooNE \cite{1303.2588},
BNL-E776 \cite{Borodovsky:1992pn},
KARMEN \cite{Armbruster:2002mp},
NOMAD \cite{Astier:2003gs},
ICARUS \cite{1307.4699}
and
OPERA \cite{1303.3953}
experiments.

\item
The
$\nua{e}$
disappearance data described in Ref.~\cite{1210.5715},
which take into account the
reactor
\cite{1101.2663,1101.2755,1106.0687}
and
Gallium
\cite{nucl-ex/0512041,Laveder:2007zz,hep-ph/0610352,0711.4222,1006.3244}
anomalies.

\item
The constraints on
$\nua{\mu}$
disappearance obtained from
the data of the
CDHSW experiment \cite{Dydak:1983zq},
from the analysis \cite{0705.0107} of
the data of
atmospheric neutrino oscillation experiments\footnote{
The IceCube data,
which could give a marginal contribution
\cite{1206.6903,1307.6824},
have not been considered
because the analysis is too complicated and subject to large uncertainties.
},
from the analysis \cite{1109.4033} of the
MINOS neutral-current data \cite{Adamson:2011ku}
and from the analysis of the
SciBooNE-MiniBooNE
neutrino \cite{Mahn:2011ea} and antineutrino \cite{Cheng:2012yy} data.

\end{enumerate}

\begin{table}[t]
\begin{center}
\begin{tabular}{c|cccc|cc|cc}
					&3+1							&3+1							&3+1							&3+1							&3+2							&3+2							&3+1+1							&3+1+1							\\
					&LOW							&HIG							&noMB							&noLSND							&LOW							&HIG							&LOW							&HIG							\\
\hline
$\chi^{2}_{\text{min}}$			&291.7		&261.8		&236.1		&278.4		&284.4		&256.4		&289.8		&259.0		\\
NDF					&256		&250		&218		&252		&252		&246		&253		&247		\\
GoF					& 6\%		&29\%		&19\%		&12\%		& 8\%		&31\%		& 6\%	&29\%	\\
\hline
$(\chi^{2}_{\text{min}})_{\text{APP}}$	&99.3		&77.0		&50.9		&91.8		&87.7		&69.8		&94.8		&75.5		\\
$(\chi^{2}_{\text{min}})_{\text{DIS}}$	&180.1		&180.1		&180.1		&180.1		&179.1		&179.1		&180.1		&180.1		\\
	$\Delta\chi^{2}_{\text{PG}}$	&12.7		&4.8		&5.1		&6.4		&17.7		&7.5		&14.9		&3.4		\\
	$\text{NDF}_{\text{PG}}$	&2		&2		&2		&2		&4		&4		&3		&3		\\
	$\text{GoF}_{\text{PG}}$	&0.2\%	& 9\%	& 8\%	& 4\%	&0.1\%	&11\%	&0.2\%	&34\%	\\
\hline
$\Delta\chi^{2}_{\text{NO}}$		&$47.5$		&$46.2$		&$47.1$		&$8.3$		&$54.8$		&$51.6$		&$49.4$	&$49.1$	\\
	$\text{NDF}_{\text{NO}}$	&$3$		&$3$		&$3$		&$3$		&$7$		&$7$		&$6$	&$6$	\\
	$n\sigma_{\text{NO}}$		&$6.3\sigma$	&$6.2\sigma$	&$6.3\sigma$	&$2.1\sigma$	&$6.0\sigma$	&$5.8\sigma$	&$5.8\sigma$	&$5.8\sigma$	\\
\end{tabular}
\end{center}
\caption{ \label{tab:chi}
\footnotesize
Results of the fit of short-baseline data
\cite{1308.5288}
taking into account all MiniBooNE data (LOW),
only the MiniBooNE data above 475 MeV (HIG),
without MiniBooNE data (noMB)
and without LSND data (noLSND)
in the
3+1,
3+2 and
3+1+1 schemes.
The first three lines give
the minimum $\chi^{2}$ ($\chi^{2}_{\text{min}}$),
the number of degrees of freedom (NDF) and
the goodness-of-fit (GoF).
The following five lines give the quantities
relevant for the appearance-disappearance (APP-DIS) parameter goodness-of-fit (PG)
\protect\cite{hep-ph/0304176}.
The last three lines give
the difference between the $\chi^{2}$ without short-baseline oscillations and $\chi^{2}_{\text{min}}$
($\Delta\chi^{2}_{\text{NO}}$),
the corresponding difference of number of degrees of freedom ($\text{NDF}_{\text{NO}}$)
and the resulting
number of $\sigma$'s ($n\sigma_{\text{NO}}$) for which the absence of oscillations is disfavored.
}
\end{table}

Table~\ref{tab:chi}
summarizes the statistical results obtained
in Ref.~\cite{1308.5288}
from global fits of the data above
in the
3+1,
3+2 and
3+1+1 schemes.
In the LOW fits
all the MiniBooNE data are considered,
including the anomalous low-energy bins,
which are omitted in the HIG fits.
There is also a 3+1-noMB fit without MiniBooNE data
and
a 3+1-noLSND fit without LSND data.

From Tab.~\ref{tab:chi},
one can see that in all fits which include the LSND data
the absence of short-baseline oscillations
is disfavored by about $6\sigma$,
because the improvement of the $\chi^2$ with short-baseline oscillations
is much larger than the number of oscillation parameters.

In all the
3+1,
3+2 and
3+1+1 schemes
the goodness-of-fit in the LOW analysis is significantly worse than that in the HIG analysis
and the appearance-disappearance parameter goodness-of-fit is much worse.
This result confirms the fact that the MiniBooNE low-energy anomaly
is incompatible with neutrino oscillations,
because it would require a small value of $\Delta{m}^2_{41}$
and a large value of $\sin^22\vartheta_{e\mu}$
\cite{1109.4033,1111.1069},
which are excluded by the data of other experiments
(see Ref.~\cite{1308.5288} for further details)\footnote{
One could fit the three anomalous MiniBooNE low-energy bins
in a 3+2 scheme \cite{1207.4765}
by considering the appearance data without the
ICARUS \cite{1307.4699}
and
OPERA \cite{1303.3953}
constraints,
but the corresponding relatively large transition probabilities are excluded
by the disappearance data.
}.
Note that the appearance-disappearance tension
in the 3+2-LOW fit is even worse than that in the 3+1-LOW fit,
since the $\Delta\chi^{2}_{\text{PG}}$ is so much larger that it cannot be compensated
by the additional degrees of freedom
(this behavior has been explained in Ref.~\cite{1302.6720}).
Therefore,
we think that it is very likely that the MiniBooNE low-energy anomaly
has an explanation which is different from neutrino oscillations
and the HIG fits are more reliable than the LOW fits.

The 3+2 mixing scheme,
was considered to be interesting in 2010
when the MiniBooNE neutrino
\cite{0812.2243}
and antineutrino
\cite{1007.1150}
data showed a CP-violating tension.
Unfortunately,
this tension reduced considerably in the final MiniBooNE data
\cite{1303.2588}
and from Tab.~\ref{tab:chi}
one can see that there is little improvement of the 3+2-HIG fit
with respect to the 3+1-HIG fit,
in spite of the four additional parameters and the additional possibility of CP violation.
Moreover,
since the p-value obtained by restricting the 3+2 scheme to 3+1
disfavors the 3+1 scheme only at
$1.2\sigma$
\cite{1308.5288},
we think that considering the larger complexity of the 3+2 scheme
is not justified by the data\footnote{
See however the somewhat different conclusions reached in Ref.~\cite{1303.3011}.
}.

The results of the
3+1+1-HIG
fit presented in Tab.~\ref{tab:chi}
show that
the appearance-disappearance parameter goodness-of-fit
is remarkably good,
with a
$\Delta\chi^{2}_{\text{PG}}$
that is smaller than those in the 3+1-HIG and 3+2-HIG fits.
However,
the $\chi^2_{\text{min}}$ in the 3+1+1-HIG is only slightly smaller than that in the
3+1-HIG fit
and the p-value obtained by restricting the 3+1+1 scheme to 3+1
disfavors the 3+1 scheme only at
$0.8\sigma$
\cite{1308.5288}.
Therefore,
there is no compelling reason to prefer the more complex 3+1+1
to the simpler 3+1 scheme.

\begin{figure*}[t]
\null
\hfill
\includegraphics*[width=0.49\linewidth]{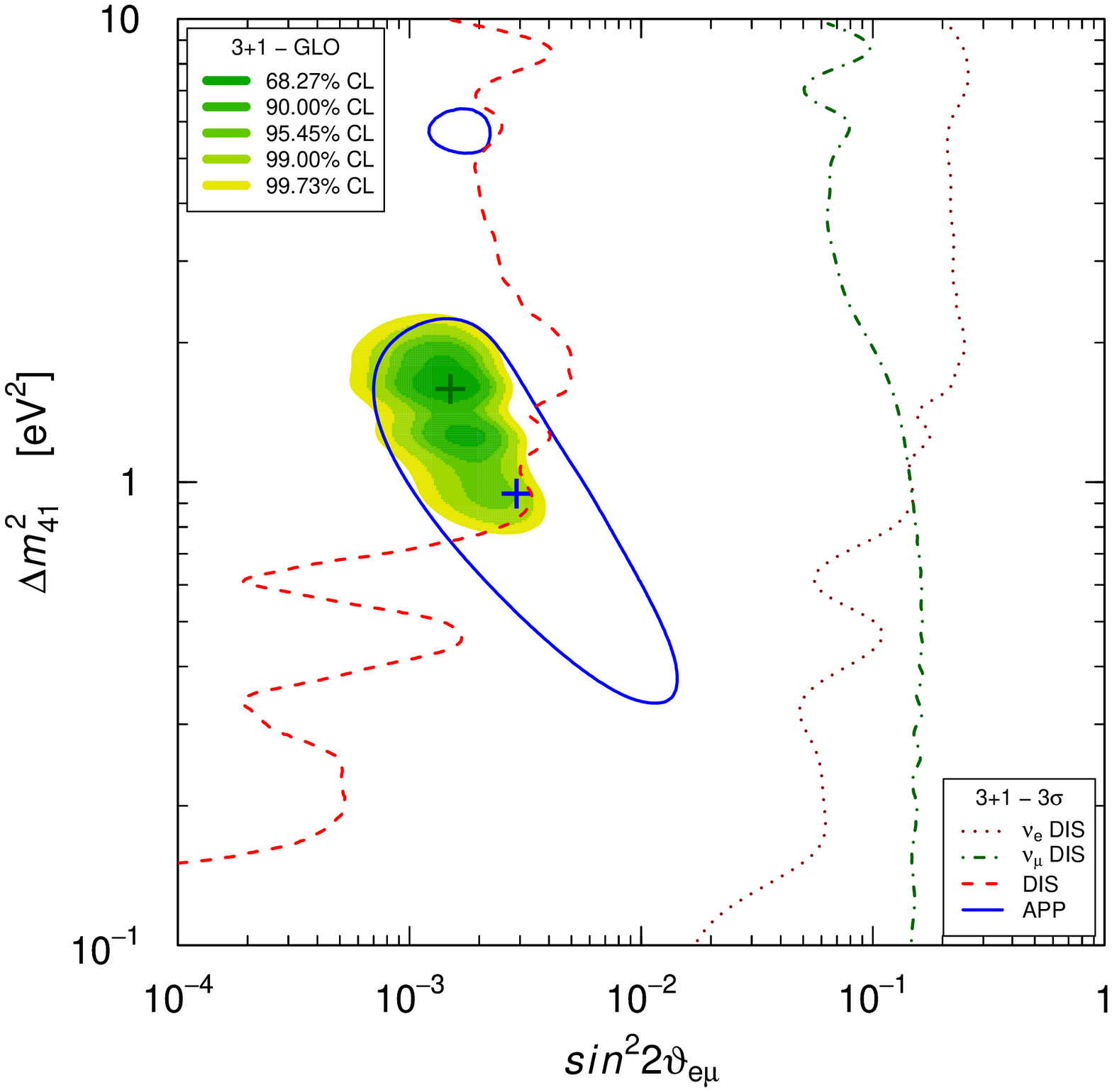}
\hfill
\includegraphics*[width=0.49\linewidth]{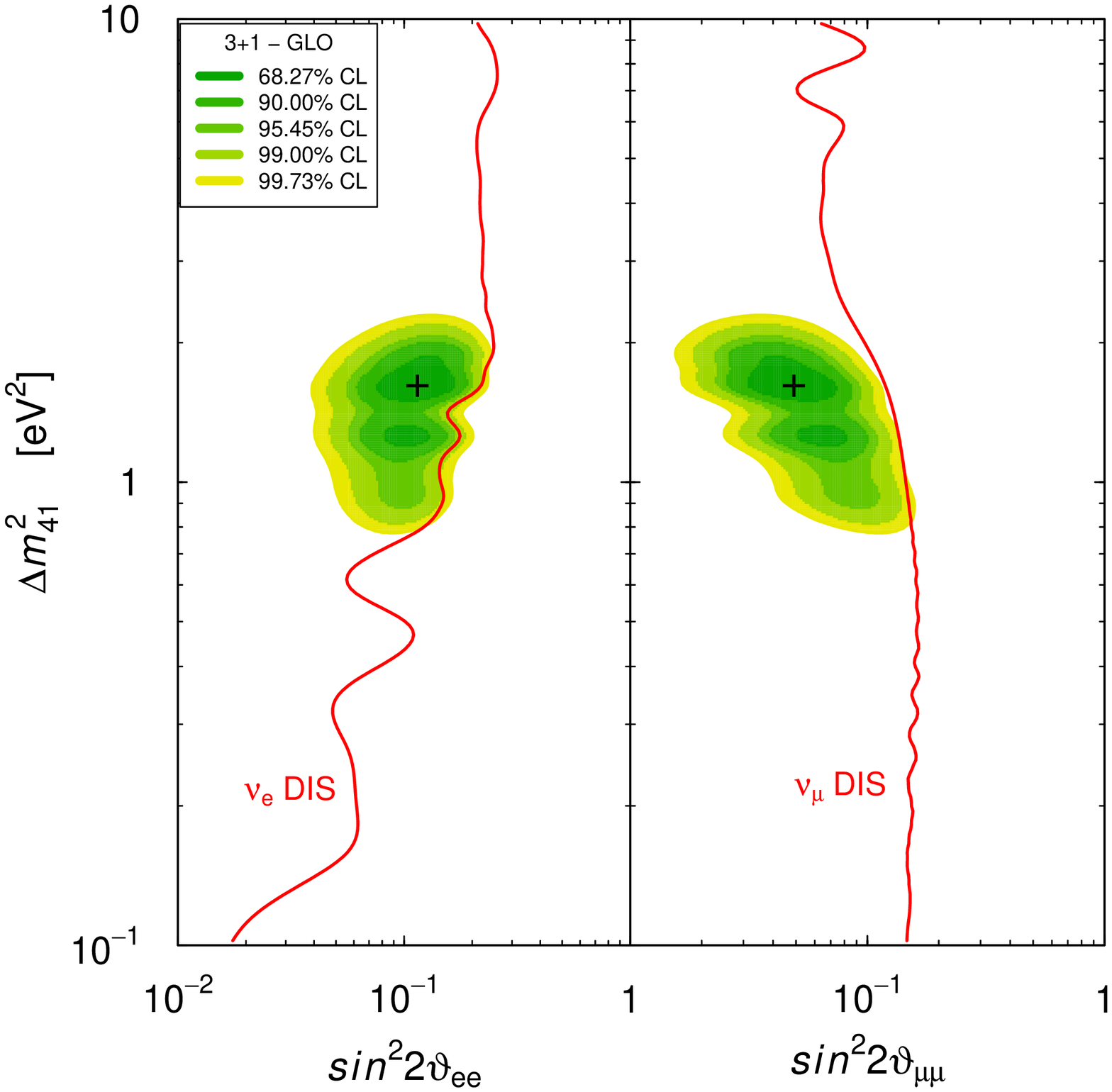}
\hfill
\null
\caption{ \label{fig:glo}
\footnotesize
Allowed regions in the
$\sin^{2}2\vartheta_{e\mu}$--$\Delta{m}^{2}_{41}$,
$\sin^{2}2\vartheta_{ee}$--$\Delta{m}^{2}_{41}$
and
$\sin^{2}2\vartheta_{\mu\mu}$--$\Delta{m}^{2}_{41}$
planes
obtained in the global (GLO) 3+1-HIG fit
\cite{1308.5288}
of short-baseline neutrino oscillation data
compared with the $3\sigma$ allowed regions
obtained from
$\protect\nua{\mu}\to\protect\nua{e}$
short-baseline appearance data (APP)
and the $3\sigma$ constraints obtained from
$\protect\nua{e}$
short-baseline disappearance data ($\nu_{e}$ DIS),
$\protect\nua{\mu}$
short-baseline disappearance data ($\nu_{\mu}$ DIS)
and the
combined short-baseline disappearance data (DIS).
The best-fit points of the GLO and APP fits are indicated by crosses.
}
\end{figure*}

Figure~\ref{fig:glo}
shows the allowed regions in the
$\sin^{2}2\vartheta_{e\mu}$--$\Delta{m}^{2}_{41}$,
$\sin^{2}2\vartheta_{ee}$--$\Delta{m}^{2}_{41}$ and
$\sin^{2}2\vartheta_{\mu\mu}$--$\Delta{m}^{2}_{41}$
planes
obtained in the 3+1-HIG fit of Ref.~\cite{1308.5288}.
These regions are relevant, respectively, for
$\nua{\mu}\to\nua{e}$ appearance,
$\nua{e}$ disappearance and
$\nua{\mu}$ disappearance
searches.
The corresponding marginal allowed intervals of the oscillation parameters
are given in Tab.~\ref{tab:int}.
Figure~\ref{fig:glo}
shows also the region allowed by $\nua{\mu}\to\nua{e}$ appearance data
and
the constraints from
$\nua{e}$ disappearance and
$\nua{\mu}$ disappearance data.
One can see that the combined disappearance constraint
in the $\sin^{2}2\vartheta_{e\mu}$--$\Delta{m}^{2}_{41}$ plane
excludes a large part of the region allowed by $\nua{\mu}\to\nua{e}$ appearance data,
leading to the well-known
appearance-disappearance tension
\cite{1103.4570,1107.1452,1109.4033,1111.1069,1207.4765,1207.6515,1302.6720,1303.3011}
quantified by the parameter goodness-of-fit in Tab.~\ref{tab:chi}.

\begin{table}[t]
\begin{center}
\begin{tabular}{c|cccc}
CL
&
$\Delta{m}^2_{41}[\text{eV}^2]$
&
$\sin^22\vartheta_{e\mu}$
&
$\sin^22\vartheta_{ee}$
&
$\sin^22\vartheta_{\mu\mu}$
\\
\hline
68.27\%
&
$ 1.55 - 1.72 $
&
$ 0.0012 - 0.0018 $
&
$ 0.089 - 0.15 $
&
$ 0.036 - 0.065 $
\\
\hline
90.00\%
&
$ 1.19 - 1.91 $
&
$ 0.001 - 0.0022 $
&
$ 0.072 - 0.17 $
&
$ 0.03 - 0.085 $
\\
\hline
95.00\%
&
$ 1.15 - 1.97 $
&
$ 0.00093 - 0.0023 $
&
$ 0.066 - 0.18 $
&
$ 0.028 - 0.095 $
\\
\hline
95.45\%
&
$ 1.14 - 1.97 $
&
$ 0.00091 - 0.0024 $
&
$ 0.065 - 0.18 $
&
$ 0.027 - 0.095 $
\\
\hline
99.00\%
&
$ 0.87 - 2.09 $
&
$ 0.00078 - 0.003 $
&
$ 0.054 - 0.2 $
&
$ 0.022 - 0.12 $
\\
\hline
99.73\%
&
$ 0.82 - 2.19 $
&
$ 0.00066 - 0.0034 $
&
$ 0.047 - 0.22 $
&
$ 0.019 - 0.14 $
\end{tabular}
\end{center}
\caption{ \label{tab:int}
\footnotesize
Marginal allowed intervals of the oscillation parameters
obtained in the global 3+1-HIG fit
of short-baseline neutrino oscillation data
\cite{1308.5288}.
}
\end{table}

It is interesting to investigate what is the
impact of the MiniBooNE experiment
on the global analysis of short-baseline neutrino oscillation data.
With this aim,
the authors of Ref.~\cite{1308.5288}
performed two additional 3+1 fits:
a 3+1-noMB fit without MiniBooNE data
and
a 3+1-noLSND fit without LSND data.
From Tab.~\ref{tab:chi}
one can see that the results of the
3+1-noMB fit are similar to those of the
3+1-HIG fit
and the exclusion of the case of no-oscillations remains at the level of $6\sigma$.
On the other hand,
in the 3+1-noLSND fit,
without LSND data,
the exclusion of the case of no-oscillations drops dramatically to
$2.1\sigma$.
In fact,
in this case
the main indication in favor of short-baseline oscillations
is given by the reactor
and
Gallium
anomalies
which have a similar statistical significance
(see Section~\ref{Introduction}).
Therefore,
it is clear that the LSND experiment is still crucial for the indication in favor of short-baseline
$\bar\nu_{\mu}\to\bar\nu_{e}$
transitions
and the MiniBooNE experiment has been rather inconclusive.

\section{Conclusions}
\label{Conclusions}

The results of the global fit of
short-baseline neutrino oscillation data presented
in Ref.~\cite{1308.5288}
show that the data can be explained by 3+1 neutrino mixing
and this simplest scheme beyond three-neutrino mixing
cannot be rejected in favor of
the more complex 3+2 and 3+1+1 schemes.
The low-energy MiniBooNE anomaly cannot be explained by neutrino oscillations
in any of these schemes.
Moreover,
the crucial indication
in favor of short-baseline
$\bar\nu_{\mu}\to\bar\nu_{e}$
appearance is still given by the old LSND data
and the MiniBooNE experiment has been inconclusive.
Hence new better experiments are needed in order to
check this signal
\cite{1204.5379,1304.2047,1307.7097,1308.0494,1308.6822}.

%\bibliographystyle{iopart-num}
%\bibliography{bibtex/nu}

\section*{References}


\begin{thebibliography}{10}
\expandafter\ifx\csname url\endcsname\relax
\def\url#1{{\tt #1}}\fi
\expandafter\ifx\csname urlprefix\endcsname\relax\def\urlprefix{URL }\fi
\providecommand{\eprint}[2][]{\url{#2}}

\bibitem{1205.4018}
Forero D, Tortola M and Valle J 2012 {\em Phys. Rev.\/} {\bf D86} 073012
(\textit{Preprint} \eprint{arXiv:1205.4018})

\bibitem{1205.5254}
Fogli G {\em et~al.\/} 2012 {\em Phys. Rev.\/} {\bf D86} 013012
(\textit{Preprint} \eprint{arXiv:1205.5254})

\bibitem{1209.3023}
Gonzalez-Garcia M, Maltoni M, Salvado J and Schwetz T 2012 {\em JHEP\/} {\bf
12} 123 (\textit{Preprint} \eprint{arXiv:1209.3023})

\bibitem{Giunti-Kim-2007}
Giunti C and Kim C~W 2007 {\em {Fundamentals of Neutrino Physics and
Astrophysics}\/} (Oxford, UK: Oxford University Press) {ISBN
978-0-19-850871-7}

\bibitem{0704.1800}
Gonzalez-Garcia M~C and Maltoni M 2008 {\em Phys. Rept.\/} {\bf 460} 1--129
(\textit{Preprint} \eprint{arXiv:0704.1800})

\bibitem{nucl-ex/9504002}
Athanassopoulos C {\em et~al.\/} (LSND) 1995 {\em Phys. Rev. Lett.\/} {\bf 75}
2650--2653 (\textit{Preprint} \eprint{nucl-ex/9504002})

\bibitem{hep-ex/0104049}
Aguilar A {\em et~al.\/} (LSND) 2001 {\em Phys. Rev.\/} {\bf D64} 112007
(\textit{Preprint} \eprint{hep-ex/0104049})

\bibitem{1101.2755}
Mention G {\em et~al.\/} 2011 {\em Phys. Rev.\/} {\bf D83} 073006
(\textit{Preprint} \eprint{arXiv:1101.2755})

\bibitem{1101.2663}
Mueller T~A {\em et~al.\/} 2011 {\em Phys. Rev.\/} {\bf C83} 054615
(\textit{Preprint} \eprint{arXiv:1101.2663})

\bibitem{1106.0687}
Huber P 2011 {\em Phys. Rev.\/} {\bf C84} 024617 (\textit{Preprint}
\eprint{arXiv:1106.0687})

\bibitem{nucl-ex/0512041}
Abdurashitov J~N {\em et~al.\/} (SAGE) 2006 {\em Phys. Rev.\/} {\bf C73} 045805
(\textit{Preprint} \eprint{nucl-ex/0512041})

\bibitem{Laveder:2007zz}
Laveder M 2007 {\em Nucl. Phys. Proc. Suppl.\/} {\bf 168} 344--346 {Workshop on
Neutrino Oscillation Physics (NOW 2006), Otranto, Lecce, Italy, 9-16 Sep
2006}

\bibitem{hep-ph/0610352}
Giunti C and Laveder M 2007 {\em Mod. Phys. Lett.\/} {\bf A22} 2499--2509
(\textit{Preprint} \eprint{hep-ph/0610352})

\bibitem{1006.3244}
Giunti C and Laveder M 2011 {\em Phys. Rev.\/} {\bf C83} 065504
(\textit{Preprint} \eprint{arXiv:1006.3244})

\bibitem{1210.5715}
Giunti C, Laveder M, Li Y, Liu Q and Long H 2012 {\em Phys. Rev.\/} {\bf D86}
113014 (\textit{Preprint} \eprint{arXiv:1210.5715})

\bibitem{1001.2731}
Kaether F, Hampel W, Heusser G, Kiko J and Kirsten T 2010 {\em Phys. Lett.\/}
{\bf B685} 47--54 (\textit{Preprint} \eprint{arXiv:1001.2731})

\bibitem{0901.2200}
Abdurashitov J~N {\em et~al.\/} (SAGE) 2009 {\em Phys. Rev.\/} {\bf C80} 015807
(\textit{Preprint} \eprint{arXiv:0901.2200})

\bibitem{hep-ph/9606411}
Okada N and Yasuda O 1997 {\em Int. J. Mod. Phys.\/} {\bf A12} 3669--3694
(\textit{Preprint} \eprint{hep-ph/9606411})

\bibitem{hep-ph/9607372}
Bilenky S~M, Giunti C and Grimus W 1998 {\em Eur. Phys. J.\/} {\bf C1} 247--253
(\textit{Preprint} \eprint{hep-ph/9607372})

\bibitem{hep-ph/9903454}
Bilenky S~M, Giunti C, Grimus W and Schwetz T 1999 {\em Phys. Rev.\/} {\bf D60}
073007 (\textit{Preprint} \eprint{hep-ph/9903454})

\bibitem{hep-ph/0405172}
Maltoni M, Schwetz T, Tortola M and Valle J 2004 {\em New J. Phys.\/} {\bf 6}
122 (\textit{Preprint} \eprint{hep-ph/0405172})

\bibitem{hep-ph/0305255}
Sorel M, Conrad J and Shaevitz M 2004 {\em Phys. Rev.\/} {\bf D70} 073004
(\textit{Preprint} \eprint{hep-ph/0305255})

\bibitem{hep-ph/0609177}
Karagiorgi G {\em et~al.\/} 2007 {\em Phys. Rev.\/} {\bf D75} 013011
(\textit{Preprint} \eprint{hep-ph/0609177})

\bibitem{0705.0107}
Maltoni M and Schwetz T 2007 {\em Phys. Rev.\/} {\bf D76} 093005
(\textit{Preprint} \eprint{arXiv:0705.0107})

\bibitem{0906.1997}
Karagiorgi G, Djurcic Z, Conrad J, Shaevitz M~H and Sorel M 2009 {\em Phys.
Rev.\/} {\bf D80} 073001 (\textit{Preprint} \eprint{arXiv:0906.1997})

\bibitem{1010.3970}
Nelson A~E 2011 {\em Phys. Rev.\/} {\bf D84} 053001 (\textit{Preprint}
\eprint{arXiv:1010.3970})

\bibitem{1201.6662}
Fan J and Langacker P 2012 {\em JHEP\/} {\bf 04} 083 (\textit{Preprint}
\eprint{arXiv:1201.6662})

\bibitem{1205.1791}
Kuflik E, McDermott S~D and Zurek K~M 2012 {\em Phys. Rev.\/} {\bf D86} 033015
(\textit{Preprint} \eprint{arXiv:1205.1791})

\bibitem{1306.6079}
Huang J and Nelson A~E 2013 {\em Phys.Rev.\/} {\bf D88} 033016
(\textit{Preprint} \eprint{arXiv:1306.6079})

\bibitem{Pontecorvo:1968fh}
Pontecorvo B 1968 {\em Sov. Phys. JETP\/} {\bf 26} 984--988

\bibitem{hep-ph/0111326}
Volkas R~R 2002 {\em Prog. Part. Nucl. Phys.\/} {\bf 48} 161--174
(\textit{Preprint} \eprint{hep-ph/0111326})

\bibitem{hep-ph/0603118}
Mohapatra R~N and Smirnov A~Y 2006 {\em Ann. Rev. Nucl. Part. Sci.\/} {\bf 56}
569--628 (\textit{Preprint} \eprint{hep-ph/0603118})

\bibitem{1206.6231}
Diaferio A and Angus G~W 2012 (\textit{Preprint} \eprint{arXiv:1206.6231})

\bibitem{1301.7102}
Riemer-Sorensen S, Parkinson D and Davis T~M 2013 (\textit{Preprint}
\eprint{arXiv:1301.7102})

\bibitem{1307.0637}
Archidiacono M, Giusarma E, Hannestad S and Mena O 2013 (\textit{Preprint}
\eprint{arXiv:1307.0637})

\bibitem{1303.3011}
Kopp J, Machado P~A~N, Maltoni M and Schwetz T 2013 {\em JHEP\/} {\bf 1305} 050
(\textit{Preprint} \eprint{arXiv:1303.3011})

\bibitem{1308.5288}
Giunti C, Laveder M, Li Y and Long H 2013 (\textit{Preprint}
\eprint{arXiv:1308.5288})

\bibitem{0812.2243}
Aguilar-Arevalo A~A {\em et~al.\/} (MiniBooNE) 2009 {\em Phys. Rev. Lett.\/}
{\bf 102} 101802 (\textit{Preprint} \eprint{arXiv:0812.2243})

\bibitem{1007.1150}
Aguilar-Arevalo A~A {\em et~al.\/} (MiniBooNE) 2010 {\em Phys. Rev. Lett.\/}
{\bf 105} 181801 (\textit{Preprint} \eprint{arXiv:1007.1150})

\bibitem{1303.2588}
Aguilar-Arevalo A {\em et~al.\/} (MiniBooNE) 2013 {\em Phys.Rev.Lett.\/} {\bf
110} 161801 (\textit{Preprint} \eprint{arXiv:1303.2588})

\bibitem{1109.4033}
Giunti C and Laveder M 2011 {\em Phys.Rev.\/} {\bf D84} 093006
(\textit{Preprint} \eprint{arXiv:1109.4033})

\bibitem{1111.1069}
Giunti C and Laveder M 2011 {\em Phys. Lett.\/} {\bf B706} 200--207
(\textit{Preprint} \eprint{arXiv:1111.1069})

\bibitem{Martini:2012fa}
Martini M, Ericson M and Chanfray G 2012 {\em Phys. Rev.\/} {\bf D85} 093012
(\textit{Preprint} \eprint{arXiv:1202.4745})

\bibitem{Martini:2012uc}
Martini M, Ericson M and Chanfray G 2013 {\em Phys. Rev.\/} {\bf D87} 013009
(\textit{Preprint} \eprint{arXiv:1211.1523})

\bibitem{Borodovsky:1992pn}
Borodovsky L {\em et~al.\/} (BNL-E776) 1992 {\em Phys. Rev. Lett.\/} {\bf 68}
274--277

\bibitem{Armbruster:2002mp}
Armbruster B {\em et~al.\/} (KARMEN) 2002 {\em Phys. Rev.\/} {\bf D65} 112001
(\textit{Preprint} \eprint{hep-ex/0203021})

\bibitem{Astier:2003gs}
Astier P {\em et~al.\/} (NOMAD) 2003 {\em Phys. Lett.\/} {\bf B570} 19--31
(\textit{Preprint} \eprint{hep-ex/0306037})

\bibitem{1307.4699}
Antonello M {\em et~al.\/} (ICARUS) 2013 (\textit{Preprint}
\eprint{arXiv:1307.4699})

\bibitem{1303.3953}
Agafonova N {\em et~al.\/} (OPERA) 2013 {\em JHEP\/} {\bf 1307} 004
(\textit{Preprint} \eprint{arXiv:1303.3953})

\bibitem{0711.4222}
Acero M~A, Giunti C and Laveder M 2008 {\em Phys. Rev.\/} {\bf D78} 073009
(\textit{Preprint} \eprint{arXiv:0711.4222})

\bibitem{Dydak:1983zq}
Dydak F {\em et~al.\/} (CDHSW) 1984 {\em Phys. Lett.\/} {\bf B134} 281

\bibitem{1206.6903}
Esmaili A, Halzen F and Peres O~L~G 2012 {\em JCAP\/} {\bf 1211} 041
(\textit{Preprint} \eprint{arXiv:1206.6903})

\bibitem{1307.6824}
Esmaili A and Smirnov A~Y 2013 (\textit{Preprint} \eprint{arXiv:1307.6824})

\bibitem{Adamson:2011ku}
Adamson P {\em et~al.\/} (MINOS) 2011 {\em Phys. Rev. Lett.\/} {\bf 107} 011802
(\textit{Preprint} \eprint{arXiv:1104.3922})

\bibitem{Mahn:2011ea}
Mahn K~B~M {\em et~al.\/} (SciBooNE-MiniBooNE) 2012 {\em Phys. Rev.\/} {\bf
D85} 032007 (\textit{Preprint} \eprint{arXiv:1106.5685})

\bibitem{Cheng:2012yy}
Cheng G {\em et~al.\/} (SciBooNE-MiniBooNE) 2012 {\em Phys. Rev.\/} {\bf D86}
052009 (\textit{Preprint} \eprint{arXiv:1208.0322})

\bibitem{hep-ph/0304176}
Maltoni M and Schwetz T 2003 {\em Phys. Rev.\/} {\bf D68} 033020
(\textit{Preprint} \eprint{hep-ph/0304176})

\bibitem{1207.4765}
Conrad J, Ignarra C, Karagiorgi G, Shaevitz M and Spitz J 2013 {\em Adv.High
Energy Phys.\/} {\bf 2013} 163897 (\textit{Preprint}
\eprint{arXiv:1207.4765})

\bibitem{1302.6720}
Archidiacono M, Fornengo N, Giunti C, Hannestad S and Melchiorri A 2013 {\em
Phys.Rev.\/} {\bf D87} 125034 (\textit{Preprint} \eprint{arXiv:1302.6720})

\bibitem{1103.4570}
Kopp J, Maltoni M and Schwetz T 2011 {\em Phys. Rev. Lett.\/} {\bf 107} 091801
(\textit{Preprint} \eprint{arXiv:1103.4570})

\bibitem{1107.1452}
Giunti C and Laveder M 2011 {\em Phys.Rev.\/} {\bf D84} 073008
(\textit{Preprint} \eprint{arXiv:1107.1452})

\bibitem{1207.6515}
Archidiacono M, Fornengo N, Giunti C and Melchiorri A 2012 {\em Phys. Rev.\/}
{\bf D86} 065028 (\textit{Preprint} \eprint{arXiv:1207.6515})

\bibitem{1204.5379}
Abazajian K~N {\em et~al.\/} 2012 (\textit{Preprint} \eprint{arXiv:1204.5379})

\bibitem{1304.2047}
Rubbia C, Guglielmi A, Pietropaolo F and Sala P 2013 (\textit{Preprint}
\eprint{arXiv:1304.2047})

\bibitem{1307.7097}
Elnimr M {\em et~al.\/} (OscSNS) 2013 (\textit{Preprint}
\eprint{arXiv:1307.7097})

\bibitem{1308.0494}
Delahaye J~P {\em et~al.\/} 2013 (\textit{Preprint} \eprint{arXiv:1308.0494})

\bibitem{1308.6822}
Adey D {\em et~al.\/} (nuSTORM) 2013 (\textit{Preprint}
\eprint{arXiv:1308.6822})

\end{thebibliography}

\end{document}